# Phase transitions, energy storage performances and electrocaloric effect of the lead-free $Ba_{0.85}Ca_{0.15}Zr_{0.10}Ti_{0.90}O_3$ ceramic relaxor

Zouhair Hanani [a,b,*], Daoud Mezzane [a], M'barek Amjoud [a], Anna G. Razumnaya [c, d], Sébastien Fourcade [b], Yaovi Gagou [d], Khalid Hoummada [e], Mimoun El Marssi [d] and Mohamed Gouné [b]

[a] LMCN, Cadi Ayyad University, Marrakesh, 40000, Morocco
[b] ICMCB, University of Bordeaux, Pessac, 33600, France
[c] Faculty of Physics, Southern Federal University, Rostov-on-Don, 344090, Russia
[d] LPMC, University of Picardy Jules Verne, Amiens, 80039, France
[e] IM2NP, Aix Marseille University, UMR 7334, 13397, Marseille, France

[*] *Corresponding author:* e-mail: zouhair.hanani@edu.uca.ma ; Tel: +212-5 24 43 31 63

## Abstract

Lead-free $Ba_{0.85}Ca_{0.15}Zr_{0.10}Ti_{0.90}O_3$ (BCZT) ceramic exhibits excellent dielectric, ferroelectric and piezoelectric properties at the Morphotropic Phase Boundary (MPB). Previously, we demonstrated that the use of the anionic surfactant Sodium Dodecyl Sulfate (SDS, $NaC_{12}H_{25}SO_4$) could enhance the dielectric properties of BCZT ceramic using surfactant-assisted solvothermal processing [1]. In the present study, structural, dielectric, ferroelectric properties, as well as electrocaloric effect and energy storage performances of this BCZT ceramic were thoroughly investigated. X-ray diffraction (XRD) measurements revealed the presence of single perovskite phase at room temperature with the coexistence of orthorhombic and tetragonal symmetries. In-situ Raman spectroscopy results confirmed the existence of all phase transitions from rhombohedral through orthorhombic and tetragonal to cubic symmetries when the temperature varies as reported in undoped-$BaTiO_3$. Evolution of energy storage performances with temperature have been investigated. BCZT ceramic exhibits a high energy storage efficiency of ~80% at 120 °C. In addition, the electrocaloric responsivity was found to be $0.164 \times 10^{-6}$ K.m/V at 360 K.

# 1. Introduction

For several decades, lead-based perovskite ferroelectric materials such as Pb(Zr,Ti)O$_3$ (PZT) are widely used in sensors, ultrasonic generators, resonators, actuators and much more electronic devices due to their excellent ferroelectric and piezoelectric properties [2,3]. However, in view of the concern with environmental pollution and human health, lead oxide, which is a component of PZT, is highly toxic [4]. It is one of the reasons why intensive researches were conducted to find the appropriate alternative material to lead-based ceramics [5,6].

In this purpose, barium titanate (BaTiO$_3$) is usually regarded as one of potentially promising lead-free ceramics, which was the first material practically, used to fabricate piezoelectric ceramics. It is widely employed in modern technologies such as mobile electronic devices, and hybrid electric vehicles [7–9]. Furthermore, BaTiO$_3$ is also a bioceramic material which does not contain any toxic or volatile elements, and its properties can be easily tailored by versatile engineering [10–12]. However, its high dielectric constant ($\varepsilon_{max} > 10\,000$) decreases sharply to about 2000 at Curie temperature ($T_c = 120$ °C) because of the tetragonal/cubic phase transition [13]. To resolve this issue, researchers focused on doping elements to modify BaTiO$_3$ ceramics. For instance, Ca$^{2+}$ and Zr$^{4+}$ can be introduced into BaTiO$_3$ crystal lattice to replace Ba$^{2+}$ and Ti$^{4+}$, respectively [8,14]. Therefore, these process lead to an enhancement of the relaxor behavior of barium titanate materials, broadening of the maximum at Curie point, and improving the dielectric constant with peak values as high as $\varepsilon_{max} \approx 18\,000$ [15,16].

Among this family, x[Ba(Zr$_{0.2}$Ti$_{0.8}$)O$_3$]–(1-x)[(Ba$_{0.7}$Ca$_{0.3}$)TiO$_3$] (BZT-$x$BCT, BCZT) system has been reported by Liu and Ren [17] to exhibit interesting dielectric, ferroelectric, and piezoelectric properties at the Morphotropic Phase Boundary (MPB) which is comparable to those of PZT ceramics [5]. The high piezoelectric properties of this system are due to the coexistence of two ferroelectric phases at the MPB [5]. With a great interest in the MPB composition of the BZT-$x$BCT system, the synthesis and characterization of nanostructured materials provide strong motivation to understand its properties. However, the traditional techniques for preparing BCZT nanostructures rely on high pressure or high temperature [6,18]. Solvothermal synthesis provides a design of nanomaterials with specifically tailored architectures via the synthesis of nanoscale building blocks with an appropriate size and shape, and controlled orientation of the final products. In our previous work, we reported the effect of the nature of surfactant (anionic or cationic) on the morphology and the dielectric properties of BCZT ceramics [1]. We found that the anionic surfactant sodium dodecyl sulfate (SDS, NaC$_{12}$H$_{25}$SO$_4$) played a crucial role in adjusting the size and morphology, and thus enhancing the dielectric properties of BCZT ceramic [1].

More recently, literature has emerged contradictory findings about phase transitions of 50BCT-50BZT ceramic: on one hand, some works have reported the disappearance of the orthorhombic (O) phase in 50BCT-50BZT unlike in $BaTiO_3$ and suggest the presence of unusual rhombohedral-tetragonal phase at room temperature [19–21]. On the other hand, researches have demonstrated the existence of the intermediate O-phase through XRD measurements and dielectric properties [2,22,23].

The present study set out to shine new light on these debates through the examination of phase transitions in BCZT ceramic using in-situ Raman measurements. Moreover, the structural, dielectric and ferroelectric properties upon the phase transitions using a surfactant (SDS) for morphology adjusting have been also studied. The thorough analyses of P-E hysteresis as a function of temperature, conducted on the lead-free BCZT ferroelectric relaxor allowed us to highlight the variation of the electrocaloric temperature change (ΔT) around FE-PE phase transition and the energy storage performances in this compound. Since the energy storage is efficient with slim hysteresis loop, having low coercive field and high polarization, this BCZT ferroelectric relaxor will be a potential candidate for energy storage application.

## 2. Experimental

*2.1. BCZT ceramic synthesis*

BCZT pure nanocrystalline powder were obtained through SDS-assisted solvothermal synthesis as reported previously [1]. Briefly, stoichiometric amounts of barium acetate and calcium nitrate tetrahydrate were dissolved separately in glacial acetic acid and 2-ethoxyethanol, respectively. The two solutions were then mixed together. Titanium (IV) isopropoxide and zirconium n-propoxide were added to the reaction medium according to the $Ba_{0.85}Ca_{0.15}Ti_{0.90}Zr_{0.10}O_3$ (BCZT) formula. To control BCZT microstructure, a known concentration of SDS was introduced. After one hour of continuous stirring at room temperature, the obtained solution was transferred to a 30 ml Teflon-lined stainless-steel autoclave at 180 °C in an oven for 12 h. Thereafter the reaction was completed, the sealed autoclave was cooled in the air. The resulting product was collected by centrifugation at 12 000 rpm for 10 min, and washed several times with deionized water and ethanol. Then, the final product was dried at 100 °C for 12 h and calcined at 1000 °C for 4 h to get fine BCZT powder. Finally, the obtained powder was uni-axially pressed into pellets and sintered at 1250 °C for 10h.

*2.2. Characterization*

The crystalline structure of BCZT ceramic was performed by X-ray diffraction (XRD, Panalytical X-Pert Pro). The measurement were done at room temperature under a step angle of 0.02° in the 2θ range from 10 to 80° using a Cu-K$_α$ radiation (λ ~ 1.5406 Å). Raman spectra were excited using the polarized radiation of an argon laser (λ=514.5 nm) and recorded using a Renishaw inVia Reflex spectrometer equipped with a NExT (Near-Excitation Tunable) filter for the analysis of the low-frequency spectral range down to 10 cm$^{-1}$. An optical microscope with a 50× objective was used to focus the incident light as a 2-μm-diameter spot on the sample. For temperature-dependent micro-Raman measurements, we used Linkam heating and freezing stage with temperature stability of 0.1K. The microstructure of BCZT ceramic was analyzed using a Scanning Electron Microscope (SEM, Tescan VEGA-3). A precision LCR Meter (HP 4284A, 20 Hz to 1 MHz) was used to measure the electrical properties of BCZT pellets electroded with silver paste in the frequency range 20 Hz to 1 MHz. The ferroelectric hysteresis loops were measured using a ferroelectric test system (AiXACCT, TF Analyzer 1000). The electrocaloric adiabatic temperature variation and responsivity were calculated in the so called indirect method from registered P-E hysteresis loops at 1 Hz as a function of temperature.

## 3. Results and discussion

*3.1. Structural study*

Fig. 1a shows XRD pattern of BCZT ceramic at room temperature that exhibits pure perovskite phase, without any secondary phase. The peaks splitting observed around 45° (Fig. 1b) and the formation of triplet at 65.8° (Fig. 1c) indicate the formation of BCZT ceramic at the Morphotropic Phase Boundary (MPB) with the coexistence of orthorhombic (Amm2) and tetragonal phases (P4mm) [1,24–26].

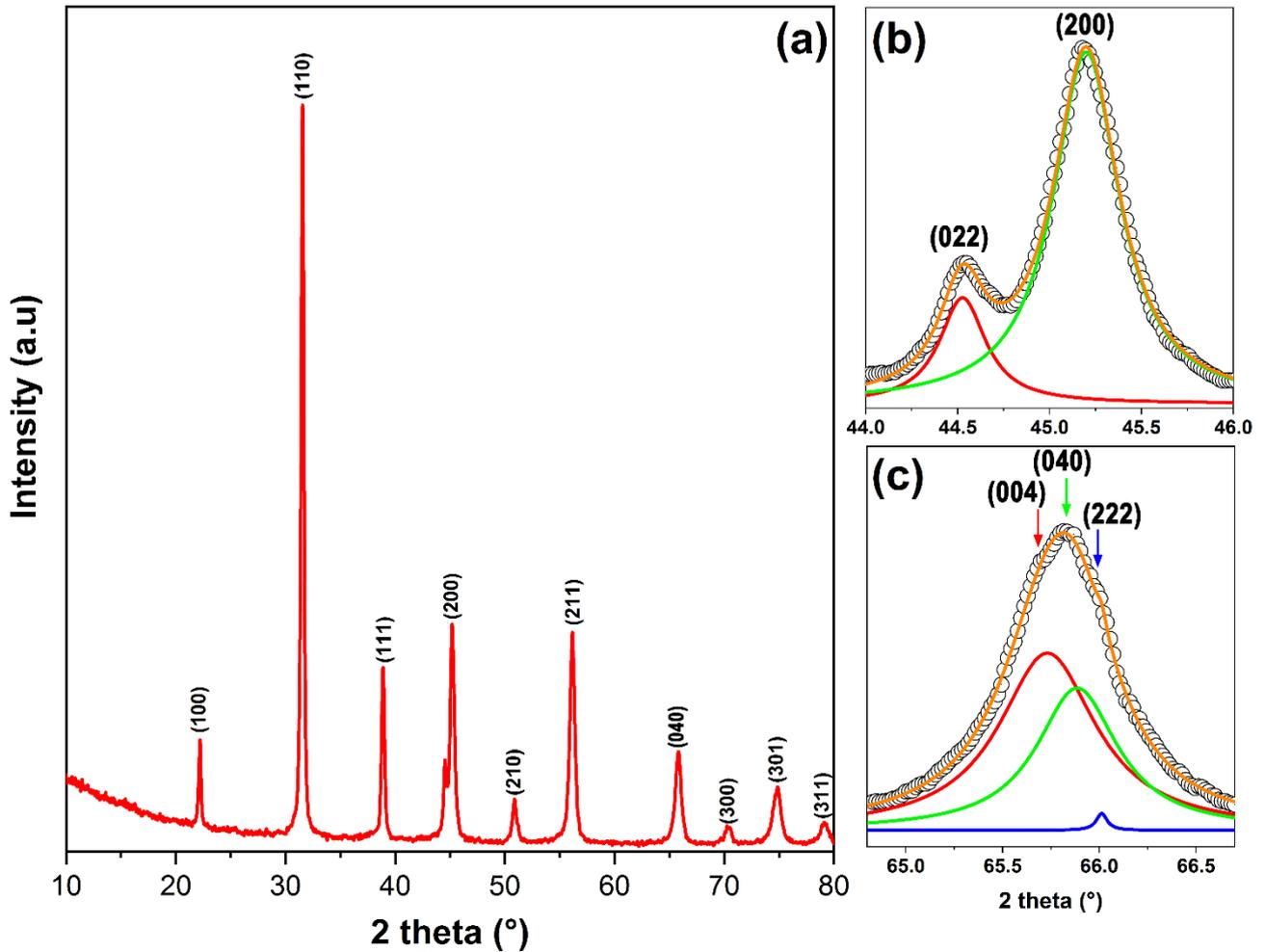

**Fig. 1.** (a) Room-temperature XRD pattern, (b) enlarged peaks splitting around 2θ ≈45° and (c) around 2θ ≈ 65.8°.

Raman spectroscopy is a powerful tool to divulge the phase transitions of complex ferroelectric materials. Fig. 2a shows Raman spectra of BCZT ceramic recorded from -30 to 140 °C. Raman peaks at around 298 and 725 cm$^{-1}$ are characteristic to the ferroelectric phases (Rhombohedral, Orthorhombic and Tetragonal) of BCZT ceramic. The Rhombohedral structure was identified through the existence of the triple modes $A_1(TO_2)$, $A_1(TO_3)$ and $E(TO_1)$ at low temperature [27]. The cubic paraelectric phase is characterized by two very broad peaks centered at about 285 and 520 cm$^{-1}$. The ferroelectric (Tetragonal) - paraelectric (Cubic) phase transition occurs at 90 °C and was evidenced by the disappearance of the peaks around 170 and 298 cm$^{-1}$ and the attenuation of the peak at 725 cm$^{-1}$. The $A_1g$ octahedral breathing mode at 800 cm$^{-1}$ which is independent of temperature is irrefutably chemical in nature and does not relate to structural distortions of the lattice away from the parent cubic symmetry [16]. In other words, the presence of this breathing mode is due to the presence of several dissimilar atoms at A-sites and B-sites forming a complex perovskite solid solution [28].

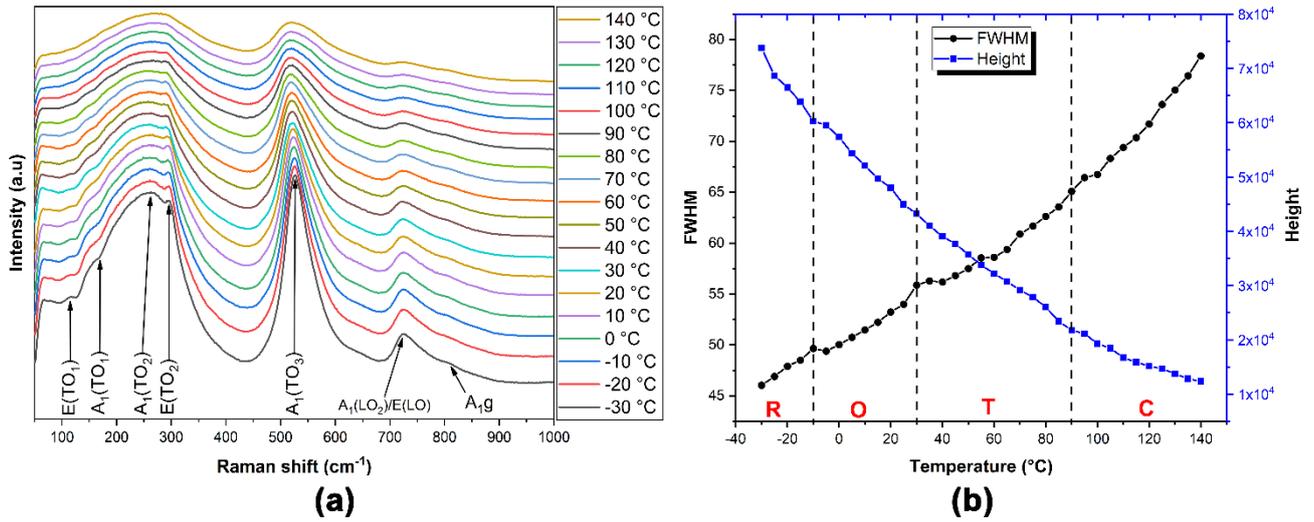

**Fig. 2.** (a) Temperature dependence of Raman spectra and (b) FWHM and peak height of the $A_1(TO_3)$ Raman mode of BCZT ceramic.

To have more insights on the phase transitions in BCZT ceramic, we attract attention to the observed broadening of the main bands in the spectral region 250 - 800 cm$^{-1}$ which is interpreted as being activated by the presence of compositional disorder [16]. Fig. 2(b) depicts the thermal variation of FWHM and peak height of $A_1(TO_3)$ Raman mode as a function of temperature. The gradual change in peak width is completely consistent with an increase in correlation length of the polar order as the temperature decreases. An abrupt variation of FWHM around -10 °C indicates the Rhombohedral (R)/Orthorhombic (O) phase transition. The second anomaly at 30 °C is consisting to Orthorhombic (O)/Tetragonal (T). The ferroelectric/paraelectric (Tetragonal (T)/Cubic (C)) transition occurred around 90 °C.

*3.2 Dielectric properties*

Fig. 3a-b display the temperature dependence of the dielectric constant ($\varepsilon_r$) and the dielectric loss (tan δ) of BCZT ceramic at different frequencies from room temperature to 200 °C. These curves show two obvious anomalies: the first around 30 °C corresponds to orthorhombic-tetragonal (O–T) transition and a second broad peak around 90 °C is associated to the tetragonal-cubic (T–C) phase transition [1,29,30]. These results bear out those obtained by Raman measurements, assuming the O–T and T–C transitions around 30 °C and 90 °C, respectively.

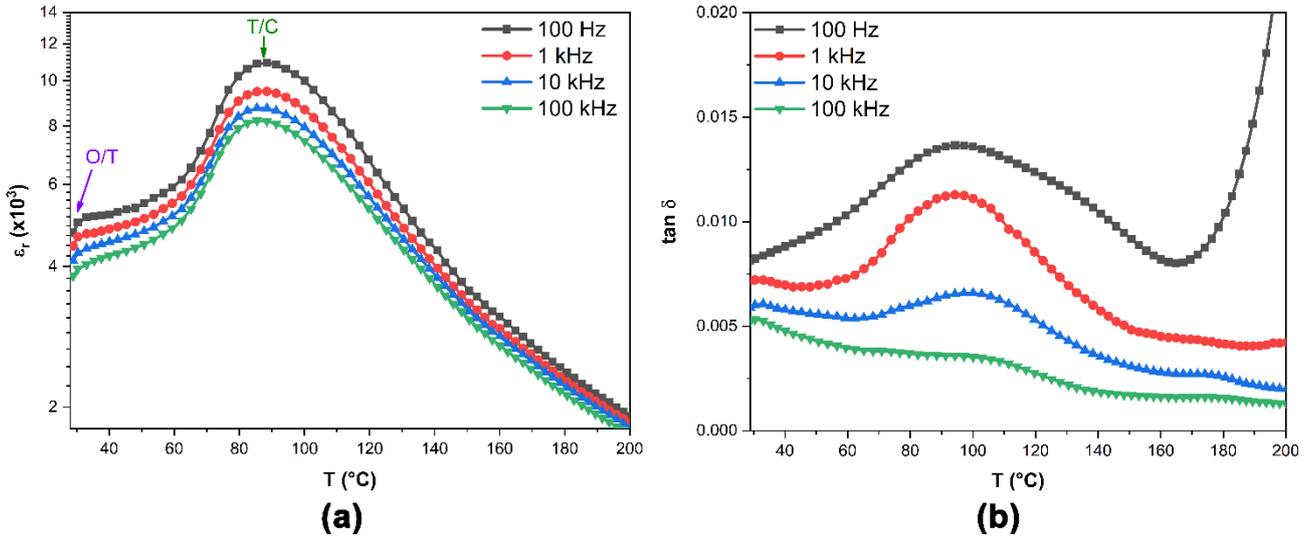

**Fig. 3.** Temperature-dependence of the (a) dielectric constant and (b) dielectric loss of BCZT ceramic at different frequencies.

*3.3 Ferroelectric properties*

Fig. 4a shows the room-temperature polarization hysteresis loop (P-E) of the BCZT ceramic obtained at different frequencies. The polarization decreases when the frequency increases in accordance with shorter time constant for dipoles ordering.

The polarization versus electric field (P-E) hysteresis loops of the BCZT ceramic registered at different temperatures at the driving frequency of 1 Hz are plotted in Fig. 4b. It presents a remnant polarization $P_r$ of ~ 2.03 $\mu C/cm^2$ and a coercive field $E_c$ of ~ 2.17 kV/cm at 0 °C. As the temperature increases, the P-E loops become slimmer accompanied by the continuous decrease of remnant polarization ($P_r$), coercive field ($E_c$) and maximal polarization ($P_{max}$) (Fig. 4c), this is due to the disappearance of ferroelectric domains. Above Curie temperature ($T_c$ = 90 °C), the P-E loops do not show strictly linear behavior, characteristic of pure paraelectric phase. However, slimmest loops appeared from Curie temperature and correspond to the existence of ferroelectric clusters or residual polar nanoregions [31–33], often attributed to the complex perovskite nature or relaxor behavior [34]. It is worth noting the presence of an anomaly at room temperature, which could be attributed to Orthorhombic/Tetragonal phase transition basing on Raman observations. Thus, the temperature dependence of the remnant polarization may indicate a change in the phase state of the material.

The current-electric field (I-E) curves of BCZT ceramic at different temperatures are presented in Fig. 4d. When a ferroelectric material is under an electric field, two types of current signals are observed, leakage current and current due to domain switching phenomena. The appearance of a peak in the current signal before reaching the maximum electric field indicates that domain switching is taking place (Fig. 4d) [35]. The magnitude of leakage current at the maximum applied electric field is very low compared with the magnitude of peak current caused by domain switching [35]. As the temperature increases, gradual changes with a broad and flat current platform are obtained in current peaks, in which the ferroelectric order was disturbed, leading to a ferroelectric-to-relaxor state transformation [36,37].

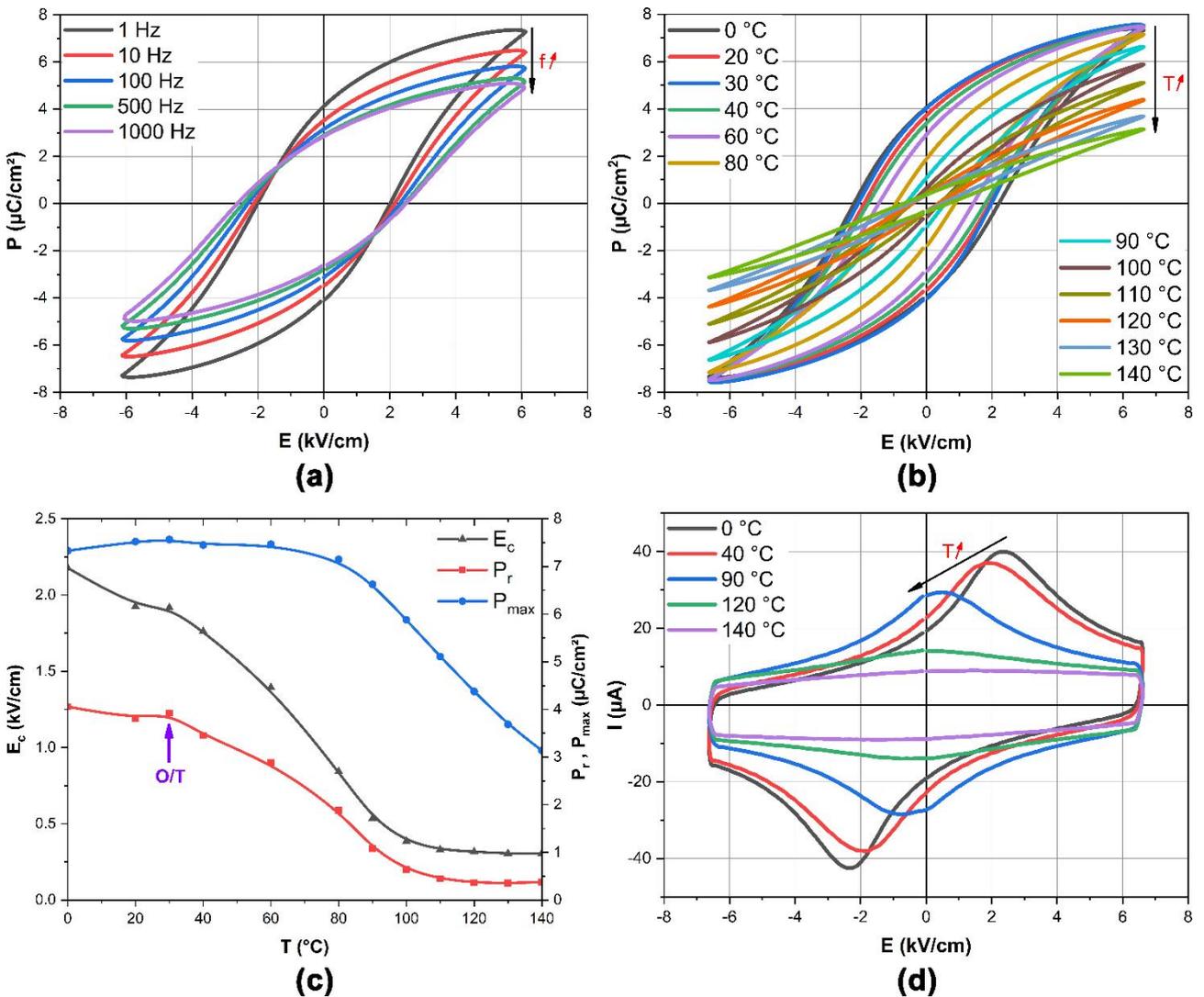

**Fig. 4.** (a) Frequency-dependence of P-E loops at room temperature, (b) temperature-dependence of P-E loops at 1 Hz, (c) variation of $P_r$, $E_c$ and $P_{max}$ with temperature and (d) temperature-dependence of I-E curves of the BCZT ceramic.

The ferroelectric property of BCZT ceramic was also investigated through the study of the hysteresis behavior reflected in Capacitance-Voltage (C-V) characteristics. At 100 Hz, the applied voltage was swept from a positive bias to a negative bias and back again. The typical butterfly loop that confirm the ferroelectric properties is shown in Fig. 5. A C-V loop will exhibit a butterfly shape only when the material is ferroelectric, and the presence of two peaks is attributed to the ferroelectric domains switching [38,39]. Moreover, the symmetric shape of the curve around the zero bias axis, indicates that the sample contains few movable ions or charge accumulation at the ceramic/electrode interface (space charge effect) [40]. Increasing the temperature results to a narrowing of the C-V curves which indicates that the process of domains switching is faster, and the saturation occurs with low energy for the ferroelectric domain alignment [41]. The relaxor behavior of BCZT ferroelectric ceramic is demonstrated by the existence of the butterfly-like loop even at high temperatures (T ≥ 90 °C).

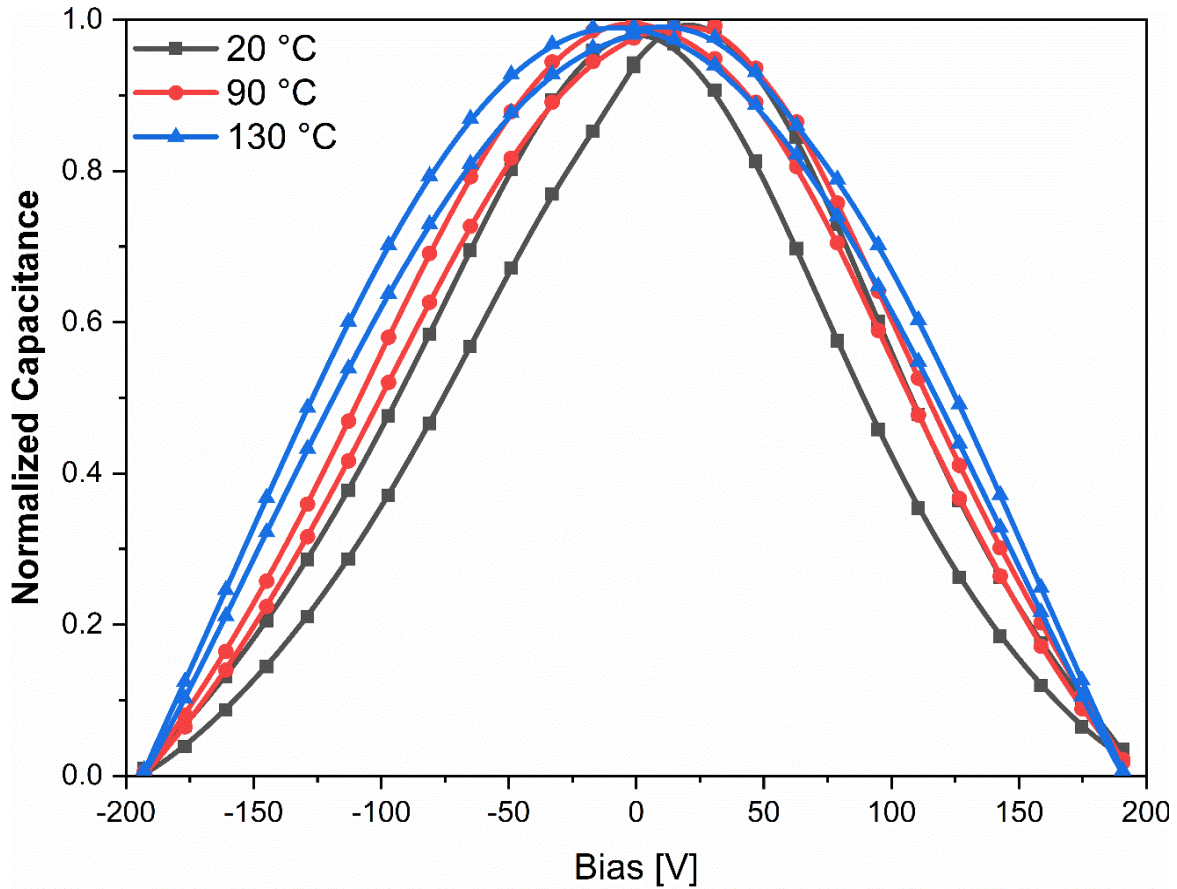

**Fig. 5.** Capacitance–Voltage (C-V) characteristic loop at 20, 90 and 130 °C at 100 Hz.

*3.4 Energy storage performances*

Previous works on energy storage properties of ceramics or thin films have been conducted mainly under very high electric field (hundreds or even thousands kV/cm). However, it is well known that a large electric field can induce higher polarization. Consequently, as both the electric field and the polarization increase, the energy-storage properties increase inevitably according to the definition of energy-storage density. Unfortunately, in relation to applications, such large electric field is not desirable for energy storage devices [42]. The key characteristic for energy storage devices is to gain a slim hysteresis loop. Relaxor ferroelectrics exhibit high permittivity and slim hysteresis loops, and consequently enhanced energy storage properties [43].

To have an insight in these performances, the recoverable and loss energy densities as well as the energy storage efficiency of the BCZT ceramic in the range of 0 – 140 °C were determined through P-E loops (Fig. 6b). The recoverable energy density ($W_{rec}$) (green area of Fig. 6a) increases gradually with temperature to reach a maximum of ~14 mJ/cm$^3$ at 90 °C, corresponding to Tetragonal/Cubic phase transition, then decreases. However, the loss energy density (W$_{loss}$) (red area of Fig. 6a) shows an inverse trend. W$_{rec}$ is relatively low as compared to literature due to the very low applied electrical field (~ 6.65 kV/cm).

For a fair comparison of the energy storage performances without the electrical field limitation, the energy storage efficiency was estimated according to the following formula (1):

$$\eta\ (\%) = \frac{W_{rec}}{W_{rec}+W_{loss}} \times 100 \qquad (1)$$

where η, W$_{rec}$ and W$_{loss}$ are the energy storage efficiency, recoverable energy density and energy loss density, respectively. From Fig. 6b, the highest energy storage efficiency is found to be ~80 % at 120 °C.

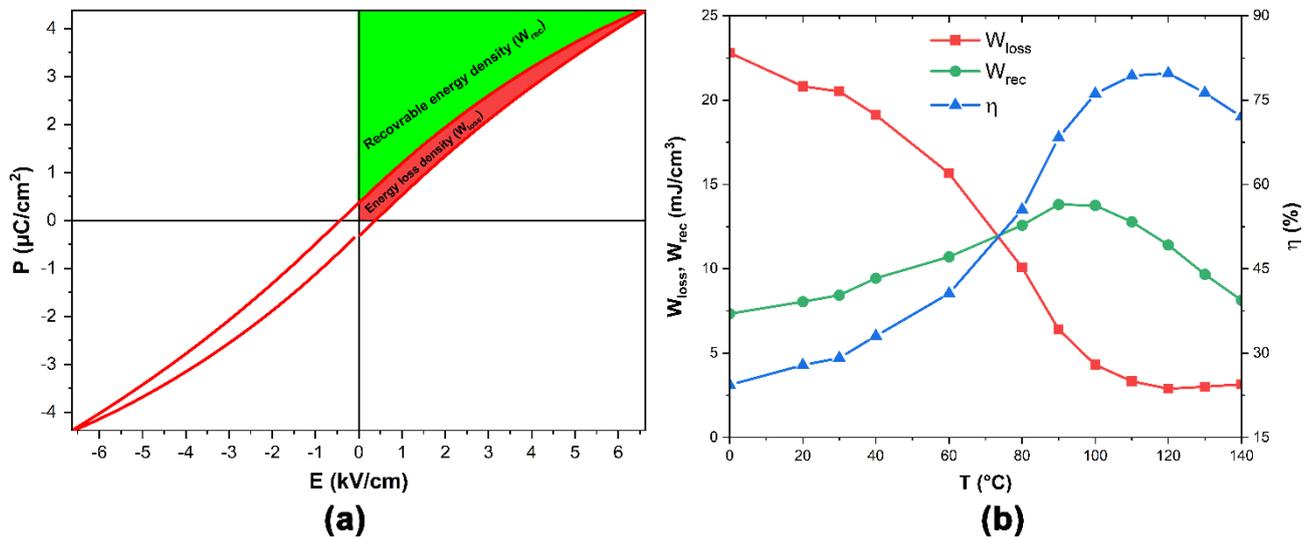

**Fig. 6.** (a) P-E loop at 120 °C with schematic calculations energy storage efficiency and (b) variation of recoverable energy density, energy loss density and the energy storage efficiency of the BCZT ceramic with temperature.

To situate our finding to literature, table 1 compares the recoverable energy density and the energy storage efficiency as a function of the applied electric field and the temperature of different lead-free ceramics.

Despite the low W$_{rec}$ due to the very low applied electrical field (~ 6.65 kV/cm), the elaborated material exhibits high energy storage efficiency (~80 %). Furthermore, the unsaturated P-E hysteresis loops indicate future opportunities to improve the energy storage capabilities of ferroelectrics [44,45]. Therefore, all these results demonstrate that the BCZT relaxor ferroelectric is a promising candidate for lead-free for energy storage application.

**Table 1.** Comparison of the energy storage performances of BCZT ceramic with other lead-free ceramics reported in literature.

| Ceramic | $W_{rec}$ (mJ/cm$^3$) | E (kV/cm) | T (°C) | η (%) | Ref. |
|---|---|---|---|---|---|
| $Ba_{0.85}Ca_{0.15}Zr_{0.10}Ti_{0.90}O_3$ (BCZT) | 14 | 6.5 | 120 | 80 | This work |
| $Ba_{0.94}Ca_{0.06}Zr_{0.16}Ti_{0.84}O_3$ | 410 | 120 | - | 72 | [46] |
| $(0.96)Bi_{0.50}(Na_{0.80}K_{0.20})_{0.50}TiO_3$-$0.04Ba_{0.90}Ca_{0.10}Ti_{0.90}Zr_{0.10}O_3$ | 373 | 45 | - | 50 | [47] |
| $(0.6)BiFeO_3$–$0.34BaTiO_3$–$0.06Ba(Zn_{1/3}Ta_{2/3})O_3$ | 1250 | 90 | 55 | 82 | [48] |
| $Na_{0.5}(Bi_{0.98}Dy_{0.02})_{0.5}TiO_3$ | 1200 | 90 | 200 | 65 | [49] |

*3.5 Electrocaloric effect*

The study of the electrocaloric effect (ECE) in the BCZT ceramic was performed to highlight the effect of elaboration technique when compares with a previously reported work on similar compounds elaborated using solid state reaction [22], and to evaluate the potential of BCZT lead-free Relaxor Ferroelectrics for environment friendly solid-state cooling devices.

The reversible adiabatic electrocaloric temperature change (ΔT) and the electrocaloric responsivity (ζ) where calculated by indirect method based on Maxwell equation, that leads to the equation (2).

$$\Delta T = -\frac{1}{\rho C_p} \int_{E_1}^{E_2} T \left(\frac{\partial P}{\partial T}\right)_E dE \qquad (2)$$

where ρ and $C_p$ are respectively, the mass density and the specific heat of the material.

After recording P-E hysteresis loops as a function of temperature, we perform a five order polynomial fitting by considering only the upper branches of these curves (Fig. 4b). Then, the variation of the polarization versus temperature at every fixed applied electric fields can be deduced (Fig. 7a). The polarization decreases rapidly with temperature around $T_c$.

Fig. 7b shows the temperature-dependence of electrocaloric adiabatic temperature change (ΔT). Each curve corresponds to a fixed applied electric field. Furthermore, these curves evidenced the ferroelectric-paraelectric phase transition around 360 K. The highest value obtained for ΔT was 0.109 K at 6.65 kV/cm at the FE-PE transition. It is worth noting that a further increase in the applied external electric field could enhance the ΔT value of the material. Therefore, the electrocaloric responsivity is much more suitable to evaluate the electrocaloric effect of materials. The obtained ΔT value corresponds to the electrocaloric responsivity ζ = 0.164 K.mm/kV.

To set out our results to literature, table 2 compares the electrocaloric properties of various lead-free ceramics. The electrocaloric responsivity value of the studied BCZT ceramic is comparable to the reported values in literature [22,50]. However, in the studied ceramic the electrocaloric responsivity is situated in mean values. This can be attributed not only to the elaboration method but also to the chemical composition, atomic rate and site occupancy in the structure.

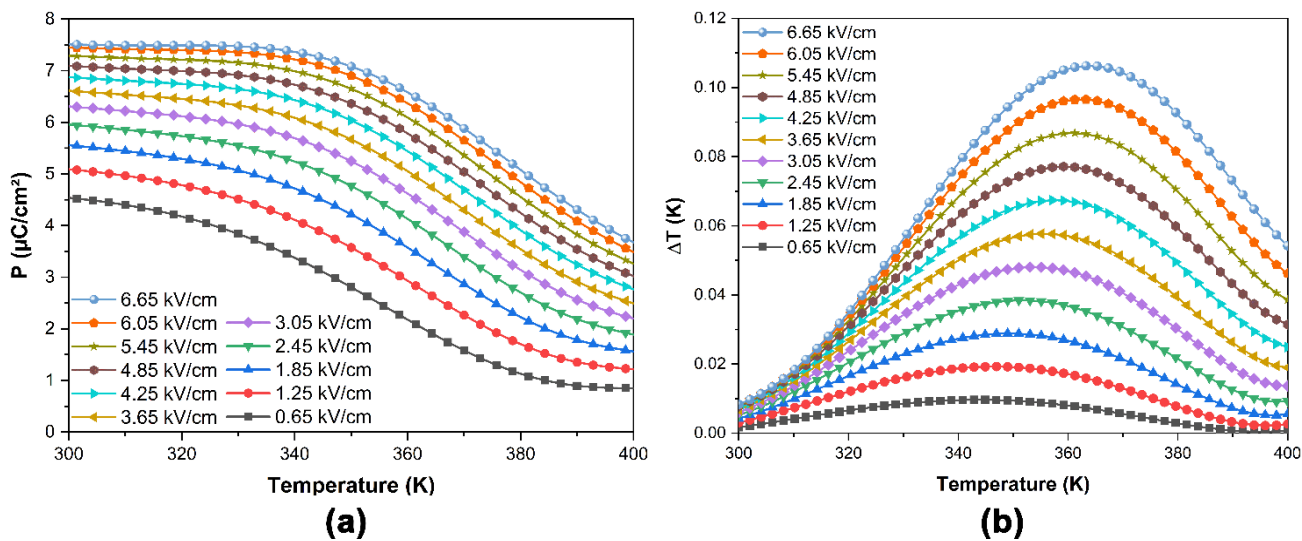

**Fig. 7.** Temperature-dependence of (a) polarization and (b) electrocaloric temperature change (ΔT) at different applied electric field.

**Table 2.** Comparison of the electrocaloric properties of BCZT ceramic with other lead-free ceramics reported in literature.

| Ceramic | $\Delta T_{max}$ (K) | $\Delta E_{max}$ (kV/cm) | T (K) | ζ (K.mm/kV) | Ref. |
|---|---|---|---|---|---|
| $Ba_{0.85}Ca_{0.15}Zr_{0.10}Ti_{0.90}O_3$ (BCZT) | 0.115 | 6.65 | 363 | 0.164 | This work |
| $Ba_{0.80}Ca_{0.20}TiO_3$ | 0.12 | 7.95 | 398 | 0.150 | [8] |
| $Ba_{0.85}Ca_{0.15}Zr_{0.10}Ti_{0.90}O_3$–Li (1wt%) | 0.328 | 20 | 346 | 0.164 | [34] |
| $Ba_{0.85}Ca_{0.15}Zr_{0.10}Ti_{0.90}O_3$ | 0.152 | 8 | 373 | 0.19 | [22] |
| $Ba_{0.80}Ca_{0.2}Zr_{0.04}Ti_{0.96}O_3$ | 0.27 | 7.95 | 386 | 0.340 | [8] |
| $Ba_{0.92}Ca_{0.08}Zr_{0.05}Ti_{0.95}O_3$ | 0.38 | 15 | 410 | 0.253 | [50] |
| $Ba_{0.91}Ca_{0.09}Zr_{0.14}Ti_{0.86}O_3$ | 0.3 | 20 | 328 | 0.150 | [51] |
| $(Ba_{1-x}Ca_x)_{1-3y/2}Bi_yTiO_3$, (x=0.05, y=0.075) | 0.81 | 39.5 | 357 | 0.207 | [52] |

## 4. Conclusions

Structural, dielectric, ferroelectric, electrocaloric and energy storage properties of the lead-free BCZT ceramic prepared by single-step surfactant-assisted solvothermal route have been successfully investigated. X-ray diffraction reveals pure perovskite phase with the coexistence of orthorhombic/tetragonal structure at room temperature. The phase transitions in BCZT ceramic were evaluated through the evolution of $\varepsilon_r$, tan $\delta$, P-E, I-E and C-V as a function of temperature and were confirmed through Raman measurements. Thus, the succession of the traditional phase transitions from rhombohedral, orthorhombic, tetragonal to cubic like undoped-$BaTiO_3$ were clearly evidenced. The maximal energy storage efficiency calculated from hysteresis loops reached ~ 80%. The electrocaloric effect was observed in BCZT ceramic, that led to the calculation of the adiabatic electrocaloric temperature change $\Delta T = 0.109$ K under 6.65 kV/cm and the electrocaloric responsivity $\zeta = 0.164$ K mm/kV at 360 °C, using indirect method. Based on the findings presented in this study, it is worth to investigate electrocaloric properties in this compound using direct method, acting on microstructure engineering of lead-free BCZT ceramics elaborated by surfactant-assisted solvothermal processing. These results demonstrate that the BCZT ferroelectric relaxor is a promising candidate for lead-free for energy storage application.

## Acknowledgements

The authors gratefully acknowledge the generous financial support of CNRST Priority Program PPR 15/2015 and the European Union's Horizon 2020 research and innovation program under the Marie Skłodowska-Curie grant agreement No 778072.

## References

1. Z. Hanani, D. Mezzane, M. Amjoud, S. Fourcade, A. G. Razumnaya, I. A. Luk'yanchuk, and M. Gouné, Superlattices Microstruct. (2018).

2. D. S. Keeble, F. Benabdallah, P. A. Thomas, M. Maglione, and J. Kreisel, Appl. Phys. Lett. **102**, 092903 (2013).

3. P. Bharathi and K. B. R. Varma, J. Appl. Phys. **116**, 164107 (2014).

4. M. Zakhozheva, L. A. Schmitt, M. Acosta, H. Guo, W. Jo, R. Schierholz, H. J. Kleebe, and X. Tan, Phys. Rev. Appl. **3**, 064018 (2015).

5. P. K. Panda and B. Sahoo, Ferroelectrics **474**, 128 (2015).

6. J. P. Praveen, T. Karthik, A. R. James, E. Chandrakala, S. Asthana, and D. Das, J. Eur. Ceram. Soc. **35**, 1785 (2015).

7. V. S. Puli, D. K. Pradhan, D. B. Chrisey, M. Tomozawa, G. L. Sharma, J. F. Scott, and R. S. Katiyar, J. Mater. Sci. **48**, 2151 (2013).

8. B. Asbani, J. L. Dellis, A. Lahmar, M. Courty, M. Amjoud, Y. Gagou, K. Djellab, D. Mezzane, Z. Kutnjak, and M. El Marssi, Appl. Phys. Lett. **106**, 042902 (2015).

9. A. Zyane, A. Belfkira, F. Brouillette, P. Marchet, and R. Lucas, J. Appl. Polym. Sci. **133**, (2016).

10. I. Coondoo, N. Panwar, H. Amorín, M. Alguero, and A. L. Kholkin, J. Appl. Phys. **113**, 214107 (2013).

11. J. Gao, X. Hu, Y. Wang, Y. Liu, L. Zhang, X. Ke, L. Zhong, H. Zhao, and X. Ren, Acta Mater. **125**, 177 (2017).

12. S. Patel, P. Sharma, and R. Vaish, Phase Transitions **89**, 1062 (2016).

13. F. Wang, W. Li, H. Jiang, M. Xue, J. Lu, and J. Yao, J. Appl. Phys. **107**, 043528 (2010).

14. B. Luo, X. Wang, Y. Wang, and L. Li, J. Mater. Chem. A **2**, 510 (2014).

15. M. A. Rafiq, M. N. Rafiq, and K. Venkata Saravanan, Ceram. Int. **41**, 11436 (2015).

16. A. Hamza, F. Benabdallah, I. Kallel, L. Seveyrat, L. Lebrun, and H. Khemakhem, J. Alloys Compd. **735**, 2523 (2018).

17. W. Liu and X. Ren, Phys. Rev. Lett. **103**, 257602 (2009).

18. B. Asbani, Y. Gagou, J. L. Dellis, A. Lahmar, M. Amjoud, D. Mezzane, Z. Kutnjak, and M. El Marssi, Solid State Commun. **237**–**238**, 49 (2016).

19. P. Jaimeewong, M. Promsawat, A. Watcharapasorn, and S. Jiansirisomboon, Integr. Ferroelectr. **175**, 25 (2016).

20. I. Coondoo, N. Panwar, R. Vidyasagar, and A. L. Kholkin, Phys. Chem. Chem. Phys. **18**, 31184 (2016).

21. H. L. Sun, Q. J. Zheng, Y. Wan, Y. Chen, X. Wu, K. W. Kwok, H. L. W. Chan, and D. M. Lin, J. Mater. Sci. Mater. Electron. **26**, 5270 (2015).

22. H. Kaddoussi, A. Lahmar, Y. Gagou, B. Manoun, J. N. Chotard, J. L. Dellis, Z. Kutnjak, H. Khemakhem, B. Elouadi, and M. El Marssi, J. Alloys Compd. **713**, 164 (2017).

23. Y. Bai, A. Matousek, P. Tofel, V. Bijalwan, B. Nan, H. Hughes, and T. W. Button, J. Eur. Ceram. Soc. **35**, 3445 (2015).

24. Z. Hanani, E.-H. Ablouh, M. Amjoud, D. Mezzane, S. Fourcade, and M. Gouné, Ceram. Int. **44**, 10997 (2018).

25. G. K. Sahoo and R. Mazumder, J. Mater. Sci. Electron. **25**, 3515 (2014).

26. D. S. Keeble, F. Benabdallah, P. A. Thomas, M. Maglione, and J. Kreisel, Appl. Phys. Lett. **102**, 092903 (2013).

27. D. A. Tenne, X. X. Xi, Y. L. Li, L. Q. Chen, A. Soukiassian, M. H. Zhu, A. R. James, J. Lettieri, D. G. Schlom, W. Tian, and X. Q. Pan, Phys. Rev. B **69**, 174101 (2004).

28. M. Ben Abdessalem, S. Aydi, A. Aydi, N. Abdelmoula, Z. Sassi, and H. Khemakhem, Appl. Phys. A **123**, 583 (2017).

29. Z. Wang, J. Wang, X. Chao, L. Wei, B. Yang, D. Wang, and Z. Yang, J. Mater. Sci. Mater. Electron. **27**, 5047 (2016).

30. W. Bai, D. Chen, J. Zhang, J. Zhong, M. Ding, B. Shen, J. Zhai, and Z. Ji, Ceram. Int. **42**, 3598 (2016).

31. G. Ramesh, M. S. Ramachandra Rao, V. Sivasubramanian, and V. Subramanian, J. Alloys Compd. **663**, 444 (2016).

32. L. Eric Cross, Ferroelectrics **76**, 241 (1987).

33. S. Tsukada, Y. Akishige, T. H. Kim, and S. Kojima, IOP Conf. Ser. Mater. Sci. Eng. **54**, 012005 (2014).

34. J. Shi, R. Zhu, X. Liu, B. Fang, N. Yuan, J. Ding, and H. Luo, Materials (Basel). **10**, 1093 (2017).

35. A. Kumar, V. V. Bhanu Prasad, K. C. James Raju, and A. R. James, Eur. Phys. J. B **88**, 287 (2015).

36. Q. Xu, Z. Song, W. Tang, H. Hao, L. Zhang, M. Appiah, M. Cao, Z. Yao, Z. He, and H. Liu, J. Am. Ceram. Soc. **98**, 3119 (2015).

37. K. Li, X. Li Zhu, X. Qiang Liu, and X. Ming Chen, Appl. Phys. Lett. **102**, 112912 (2013).

38. *Principles and Applications of Ferroelectrics and Related Materials* (n.d.).

39. H. Pei, S. Guo, L. Ren, C. Chen, B. Luo, X. Dong, K. Jin, R. Ren, and H. Muhammad Zeeshan, Sci. Rep. **7**, 6201 (2017).

40. A. Z. Simões, C. S. Riccardi, M. A. Ramírez, L. S. Cavalcante, E. Longo, and J. A. Varela,

Solid State Sci. **9**, 756 (2007).

41. G. A. Samara, J. Phys. Condens. Matter **15**, R367 (2003).

42. T. F. Zhang, X. G. Tang, Q. X. Liu, Y. P. Jiang, X. X. Huang, and Q. F. Zhou, J. Phys. D. Appl. Phys. **49**, 095302 (2016).

43. S. Tong, B. Ma, M. Narayanan, S. Liu, R. Koritala, U. Balachandran, and D. Shi, ACS Appl. Mater. Interfaces **5**, 1474 (2013).

44. H. Cheng, J. Ouyang, Y.-X. Zhang, D. Ascienzo, Y. Li, Y.-Y. Zhao, and Y. Ren, Nat. Commun. **8**, 1999 (2017).

45. Z. Diamant and W. J. Fokkens, Rhinology **39**, 187 (2001).

46. D. Zhan, Q. Xu, D. P. Huang, H. X. Liu, W. Chen, and F. Zhang, J. Phys. Chem. Solids **114**, 220 (2018).

47. D. K. Kushvaha, S. K. Rout, and B. Tiwari, J. Alloys Compd. **782**, 270 (2019).

48. N. Liu, R. Liang, Z. Zhou, and X. Dong, J. Mater. Chem. C **6**, 10211 (2018).

49. M. Benyoussef, M. Zannen, J. Belhadi, B. Manoun, J. L. Dellis, M. El Marssi, and A. Lahmar, Ceram. Int. **44**, 19451 (2018).

50. G. Singh, V. S. Tiwari, and P. K. Gupta, Appl. Phys. Lett. **103**, 202903 (2013).

51. Y. Bai, X. Han, and L. Qiao, Appl. Phys. Lett. **102**, (2013).

52. H. Zaghouene, I. Kriaa, and H. Khemakhem, Mater. Sci. Eng. B **227**, 110 (2018).